\documentclass[reprint, amsmath, amssymb, aps, superscriptaddress, nolongbibliography]{revtex4-2}

\usepackage[utf8]{inputenc}
\usepackage{hyperref,amsfonts} 

\newcommand{\EQ}[1]{\begin{align}\begin{split} #1 %Multiline equations
\end{split}\end{align}}
\newcommand{\eq}[1]{\begin{equation}{ #1 %Single line equation
}\end{equation}}

\begin{document}

\title{Notes on sums over horizons}
\author{Peng Zhao}
\affiliation{Joint School of the National University of Singapore and Tianjin University, International Campus of Tianjin University, Fuzhou 350207, China}
\author{Hong L\"u}
\affiliation{Joint School of the National University of Singapore and Tianjin University, International Campus of Tianjin University, Fuzhou 350207, China}
\affiliation{Center for Joint Quantum Studies and Department of Physics, School of Science, Tianjin University, Tianjin 300350, China}

\begin{abstract}
For a large class of black holes, we show that the sum of thermodynamic quantities over all the horizons is determined by the asymptotic data at infinity. For the Kerr-Newman metric, this proves a recent numerical observation by Hristov. We propose a new method to compute the sum of the entropies using a generalized Smarr formula. The higher-curvature corrections in Gauss-Bonnet gravity are discussed.
\end{abstract}

\maketitle

\section{Introduction and Summary}

Black holes typify the wondrous curved spacetime structures predicted by Einstein's theory of general relativity. They are asymptotic to maximally symmetric spacetimes such as Minkowski, (anti–)de Sitter [(A)dS], but with an event horizon in the middle, which is typically determined as the largest positive root of the inverse radial function. While the celebrated Schwarzschild black hole has only one such root, many other examples including the Kerr or Reissner-Nordstr\"om (RN) black holes may have two positive roots, with the smaller one corresponding to the inner horizon. In higher dimensions, and/or including the cosmological constant, complex horizons associated with the complex roots may arise. Conventionally, only the event horizon, associated with the famous Hawking radiation, has received the central attention, while the inner horizon is usually ignored and the complex horizons are treated as unphysical. Nevertheless, there are several motivations to consider all the horizons. (1) Each horizon formally satisfies the first law of black hole thermodynamics \cite{1979NCimB..51..262C}. (2) The wave equation depends on the poles at all the horizons even if one focuses only on the outer region.
(3) In the Euclidean approach to black hole thermodynamics, the metrics for rotating black holes are necessarily complex. 
(4) There are also tantalizing connections to the holography of left- and right-moving modes in conformal field theories \cite{Kraus:2006wn}.
(5) Recent progress in resolving the black hole information paradox requires the knowledge of the inside of the black hole \cite{Almheiri:2020cfm}.  
(6) For asymptotically dS spacetimes, all the horizons can be real and physically important as cosmological and black hole horizons.

Following works on the product of the entropies over all the horizons \cite{Cvetic:2010mn, Cvetic:2013eda}, the sum of the entropies was identified to obey a universal formula that depends only on the cosmological constant and the horizon topology \cite{Wang:2013smb}. This universal relation was tested for many examples and several works were dedicated to its computation \cite{Xu:2013zpa, Du:2014kpa}. Other entropy relations were explored \cite{Xu:2014qaa, Xu:2014qza}. More recently, a closely related quantity, the sum of the Euclidean actions, was found to satisfy the same universal relation. This was observed numerically by Hristov for the Kerr-Newman-AdS metric \cite{Hristov:2023cuo}. In the asymptotically flat spacetime, the sum of the Euclidean actions is a simple function of the mass and electric charge \cite{Hristov:2023sxg}.

The Euclidean action is of independent interest from the entropy as it is a quantum mechanical quantity that is related to the thermodynamic free energy via a quantum statistical relation. It is also a useful order parameter for phase transitions. The simplicity of the answer suggests that the sum over horizons is a property of the asymptotically (A)dS spacetimes, and an analytic proof would be desirable. In this note, we present an elementary proof based on the residue theorem. The main results are summarized as follows:

For a large class of black-hole solutions in Einstein gravity with a cosmological constant $\Lambda$, we prove that the sum of the inverse temperatures $\beta_h$  over all the horizons vanishes. The sum of the Euclidean actions $I_h$ is equal to the negative of the sum of the entropies. We present a new method to compute the sum of the entropies using a generalized Smarr formula that simplifies previous computations. We find
\eq{
\sum_h I_h =
\begin{cases}
\frac{(-1)^{D/2}{\mathcal A}_{D-2}}{2}\, L^{D-2} &\text{ for } D \text{ even} \\
0 &\text{ for } D \text{ odd}
\end{cases}\,.
\label{universal}
}
Here $\mathcal{A}_{D}$ is the volume of the unit $D$ sphere and $L$, the scale of the (A)dS spacetime, is related to the cosmological constant as $\Lambda = -(D-1)(D-2)/2L^2$.
Examples include the Schwarzschild, RN, and Kerr metrics in $D\ge 4$, the Kerr-Newman metric in 4D, charged, rotating black holes in $D=5,6,7$ supergravities, and and the Bañados-Teitelboim-Zanelli (BTZ) black hole in 3D.

In the asymptotically flat case, the sums are determined by the mass $M$ and electric charge $Q$ of the black hole as
\EQ{
&\sum_h \beta_h =
\begin{cases}
 8\pi M &\text{ for } D=4 \\
 0 &\text{ for } D \ne 4
\end{cases}\,, \\
&\sum_h I_h =
\begin{cases}
4 \pi M^2 - 2 \pi Q^2 &\text{ for } D=4 \\
 0 &\text{ for } D \ne 4
\end{cases}.
\label{universalflat}
}

We also discuss the higher-curvature corrections in Gauss-Bonnet gravity and obtain exact formulas for the sum of the Euclidean actions in all dimensions.

\section{The sum over horizons as a residue at infinity}
In the path-integral approach to quantum gravity, the partition function over fields that are periodic in the imaginary time is identified with the thermal partition function of the grand canonical ensemble \cite{Gibbons:1976ue}. The dominating contribution to the thermodynamic potential is given by the Euclidean action evaluated at the solutions of the classical field equations. Since the thermodynamic potential is defined at each horizon, the following relation is expected to hold in leading order:
\eq{
I(r_h) = \beta H - S - \beta \sum_i \mu_i C_i \,\Big |_{r = r_h} \,.
\label{stat}
}
Here $H$ is the enthalpy for the Gibbs free energy, $S$ is the entropy, $\mu_i$ are the chemical potentials such as the angular velocity $\Omega$ and electric potential $\Phi$, and $C_i$ are the conserved charges such as the angular momentum $J$ and the electric charge $Q$.

We denote the thermodynamic quantities at each horizon as $\beta_h \equiv \beta(r_h)$, $\Omega_h \equiv \Omega(r_h)$, etc.
The goal is to bring the inverse temperature and the chemical potentials to the form
\eq{
\beta_h = \frac{t(r_h)}{\Delta'(r_h)}\,, \quad
\beta_h \mu_{i,h}= \frac{s_i(r_h)}{\Delta'(r_h)} \,,
\label{goal}
}
for some holomorphic functions $t, s_i$ that are usually polynomials and a discriminant polynomial $\Delta$. The horizons are located at the roots of $\Delta$. We consider the case where $\Delta$ contains only simple roots. 
The sum over all the horizons may be expressed as a contour integral encircling all the simple poles. The contour may be inverted to receive a contribution only from the pole at infinity. By the residue theorem applied to a ratio of two holomorphic functions $F, G$, which has only simple poles,
\eq{
\sum_{h} \frac{F(r_h)}{G'(r_h)}
= -{\rm Res}_{z=\infty} \frac{F(z)}{G(z)} \,.
}
It follows that the residue at infinity determines the sum. If they vanish, as in spacetimes with a cosmological constant, then the sum of the Euclidean actions is reduced to the sum of the entropies.

We may prove that the inverse temperature takes this form for a large class of metrics, while the chemical potentials will be examined case by case in examples.
Consider a stationary, axisymmetric spacetime whose metric is of the form
\eq{
ds^2 = -h(r,\theta)\, dt^2 + f(r, \theta)^{-1}\, dr^2 + \cdots \,.
\label{metric}
}
$h$ may differ from $f$ but we require $h=f$ on $\theta = 0$. The horizons are defined by the roots of the metric radial function $f(r, \theta) = g_{rr}^{-1}$. We consider the case where $f(r, 0)$ contains only simple roots and no branch point. This includes the Kerr-Newman solution as well as a large class of charged, rotating black holes with scalar hairs and/or in higher dimensions.

By the zeroth law of black hole thermodynamics, the temperature is constant on the horizon. We may evaluate it at $\theta = 0$. Expanding the metric around $(r,\theta) = (r_h, 0)$ as $r = r_h + \rho^2$, the metric is conformal to $(\partial_r f(r,0))^2 \rho^2  d\tau^2 + d\rho^2$. To eliminate the deficit angle, the time coordinate must be periodic with period $t \sim t + i\beta$.
The deficit-angle argument shows that
\eq{\beta = \frac{4\pi}{\partial_r f(r,0)}\,.}
Since the zeroes of $f(r,0)$ are encoded in $\Delta$, $\beta_h$ must be of the form \eqref{goal}.

We will show that the chemical potential terms $\beta_h \mu_{i,h}$ are also of this form. Their sums over the horizons vanish in all the examples with a cosmological constant. The sum of the Euclidean actions is then equal to the negative of the sum of the entropies. The calculation of the sum of the entropies requires additional technicalities, especially when rotation is turned on. We present a simpler method using a generalized Smarr formula \cite{Kastor:2009wy}
\eq{
\frac{D-3}{D-2}M = TS + \Omega J + \frac{D-3}{D-2} \Phi Q - \frac{2}{D-2} VP  \,.
\label{Smarr}
}
For spacetimes with a cosmological constant,
$P = -\Lambda/8\pi$ is interpreted as the pressure that is conjugate to the thermodynamic volume $V$ of the black hole. Note that the black hole mass is identified with the enthalpy of the system when $\Lambda$ is allowed to vary.

The thermodynamic volume coincides with the geometric volume for static black holes but receives corrections due to rotation. If the sum of all the other thermodynamic quantities vanishes, then the sum of the entropies may be evaluated from the thermodynamic volume as
\EQ{\sum_h S_h &= \frac{D-1}{8\pi L^2} \sum_h \beta_h V_h \,,
\label{entropysum}
}
which may then be converted into a contour integral and evaluated at the pole at infinity.

\section{Examples}

\subsection{Kerr-(A)dS$_D$}
There are $N = \lfloor\frac{D-1}{2}\rfloor$ rotation variables $a_i$. The inverse temperature \cite{Hawking:1998kw, Gibbons:2004uw, Gibbons:2004ai} may be written as
\eq{
\beta(r_h) = \frac{4 \pi}{\Delta '(r_h)} \prod _{i=1}^{N} \left(r_h^2 + a_i^2\right)
\label{beta}
\,,}
where
\eq{\Delta(r) = \left(\frac{r^2}{L^2}+1\right) \prod _{i=1}^{N} \left(r^2+ a_i^2\right) - 2 m r^{2-\epsilon}\,.}
Here $\epsilon = 0$ in odd dimensions and $\epsilon = 1$ in even dimensions.
The angular velocity is
\eq{
\Omega_{i}(r_h) = \left(\frac{r_h^2}{L^2}+1\right)\frac{a_i}{r_h^2 + a_i^2} \,.
}
We see that both $\beta_h$ and $\beta_h \Omega_{i,h}$ are of the form \eqref{goal}. It follows by counting polynomial degrees that the residues at infinity vanish and $\sum_h \beta_h = \sum_h  \beta_h \Omega_{i,h} =0$.

The Bekenstein-Hawking entropy is $S_h = A_h/4$ with
\eq{
A_h = \frac{{\mathcal A}_{D-2}}{ r_h^{1-\epsilon}} \prod_{i=1}^{N} \frac{r_h^2 + a_i^2}{1-a_i^2/L^2}\,.
}
The sum of the entropies has been computed using Vieta's formulas \cite{Du:2014kpa}. 
A slick method is to apply the Smarr formula. The thermodynamic volume is \cite{Cvetic:2010jb}
\eq{
V_h = \frac{r_h}{D-1} \, A_h + \frac{8\pi}{(D-1)(D-2)}\sum_{i=1}^{N} a_i J_i \,.
}
The sum of the entropies vanishes for odd $D$ because the entropy is an odd function of $r$.
For even $D$,
\eq{
\sum_h S_h
= -\frac{{\mathcal A}_{D-2}}{2 L^2}
{\rm Res}_{r=\infty} \frac{r}{\Delta(r)} \prod_{i=1}^{N} \frac{(r^2 + a_i^2)^2}{1-a_i^2/L^2} \,,
}
and we recover the formulas \eqref{universal}.

In the flat-spacetime limit when $L\to \infty$, two horizons are sent to infinity because the degree of $\Delta$ is reduced by two. We may show that
\eq{
\sum_h \beta_h
= -{\rm Res}_{r=\infty}  \frac{4 \pi}{\Delta (r)} \prod _{i=1}^{N} \left(r^2 + a_i^2\right) \,,}
which evaluates to \eqref{universalflat}.
Similarly, $\sum_h \beta_h\Omega_h = 0$. The sum of the entropies may be evaluated by the Smarr formula and we obtain the desired result for the sum of the Euclidean actions \eqref{universalflat} with $Q=0$.

\subsection{Kerr-Newman-(A)dS}
For simplicity, we only consider a spherical horizon. The inverse temperature may be written in the same form as \eqref{beta}, where now
\eq{
\Delta(r) = \left(\frac{r^2}{L^2}+1\right)\left(r^2 + a^2\right) -2 m r+q^2\,.
\label{KNAdS4}
}
The expressions for the angular velocity and entropy are identical to Kerr-(A)dS$_4$. The electric potential is \cite{Caldarelli:1999xj}
\eq{
\Phi_h = \frac{q \, r_h}{r_h^2 + a^2}
\,.
}
The general argument implies that the sum of $\beta_h\Phi_h$ also vanishes. The sum of the entropies may be computed from the Smarr formula, or since $D=4$, directly from the coefficients of $\Delta$ using the Girard-Newton formulas, and the desired result \eqref{universal} follows.

In the flat-spacetime limit, only two horizons $r_\pm$ remain. We find from the residue at infinity that
\EQ{
\beta_+ + \beta_- &= 8\pi M \,, \\
\beta_+\Phi_+ + \beta_-\Phi_- &= 4\pi Q \,.
}
Hence the sum of the Euclidean actions follows the formula \eqref{universalflat}.

We comment on the independence of the result on the angular momentum. Using the Smarr formula \eqref{Smarr}, the Gibbs free energy may be written as
\eq{F = \frac{1}{D-2} \left( M + \Phi Q - 2VP \right)\,.}
In an asymptotically flat spacetime, the free energy and the Euclidean action do not depend on the angular momentum. 

The above methods may be extended to the $D$-dimensional RN metric straightforwardly.

\subsection{Charged, rotating black hole in $D=5$ supergravity}
This is the example considered in \cite{Chong:2005hr}. The inverse temperature is
\EQ{
&\beta_h = \frac{4 \pi}{\Delta'(r_h)} \Big[\left(r_h^2+a^2\right) \left(r_h^2+b^2\right)+a b q\Big] \,, \\
}
where
\eq{
\Delta(r) = \left(\frac{r^2}{L^2}+1\right)\left(r^2+a^2\right) \left(r^2+b^2\right)-2 m r^2+2 a b q+q^2\,.}
The other thermodynamic quantities are
\EQ{
\beta_h\Omega_{a,h} &= \frac{4\pi}{\Delta'(r_h)} \left[a \left(\frac{r_h^2}{L^2}+1\right)\left(r_h^2+b^2\right) +b q\right], \\
\beta_h \Omega_{b,h} &= \frac{4\pi}{\Delta'(r_h)} \left[b  \left(\frac{r_h^2}{L^2}+1\right)\left(r_h^2+a^2\right)+a q\right] \,, \\
\beta_h \Phi_h &= \frac{4\pi}{\Delta'(r_h)} \left(\sqrt{3}q r_h^2\right) \,, \\
S_h &= \frac{\pi^2}{2r_h} \frac{\left(r_h^2+a^2\right) \left(r_h^2+b^2\right)+a b q}{\left(1-a^2/L^2\right) \left(1-b^2/L^2\right)} \,.
}
Once written into this form, one may follow the previous arguments to show that $\sum_h \beta_h = \sum_h \beta_h \Phi_h = \sum_h  \beta_h \Omega_{i,h} =0$ in asymptotically (A)dS and flat spacetimes alike. The sum of the entropies and the sum of the Euclidean action vanish because the entropy is an odd function of $r$.

Similar calculations apply to the $D=6$ and $D=7$ supergravities considered in \cite{Chow:2008ip, Chow:2007ts}.

\subsection{BTZ black hole}
Although the BTZ metric, in its original form, does not fall into the class \eqref{metric}, it may be converted into the Kerr-AdS$_3$ form by a coordinate transform \cite{Hawking:1998kw}.
The thermodynamic quantities, in the convention of \cite{Banados:1992wn} where $8G_{\rm N} = 1$, may be written as
\EQ{
\beta_h = \frac{4\pi r_h^2}{\Delta'(r_h)} \,, \quad
\beta_h \Omega_h = \frac{2\pi J}{\Delta'(r_h)} \,,\quad
S_h = 4\pi r_h
\,,
}
where
\eq{\Delta(r) =\frac{r^4}{L^2}-M r^2 + \frac{J^2}{4}\,.
}
The Euclidean action may be directly evaluated to be
\eq{I_h = -2\pi r_h  \,, }
which is equivalent to a Smarr-type formula
\eq{M = \frac{1}{2} TS + \Omega J\,.}

\subsection{Gauss-Bonnet black hole}

It is known that higher-curvature terms lead to corrections to a seemingly universal quantity in Einstein gravity. Here we study the Gauss-Bonnet black hole with a spherical horizon \cite{Boulware:1985wk}. The inverse temperature \cite{Cai:2001dz} may be rewritten in our form as
\eq{
\beta_h = \frac{4\pi }{\Delta'(r_h)} r_h^{D-5}(r_h^2+2\tilde \alpha)\,,}
Here $\tilde \alpha$ is the Gauss-Bonnet coupling. $\Delta$ does not appear in the metric radial function but is given by
\eq{
\Delta(r) = \frac{r^{D-1}}{L^2} + r^{D-3} + \tilde \alpha  r^{D-5} -2m \,.
\label{DeltaGB}
}
The entropy is
\eq{
S_h = \frac{\mathcal{A}_{D-2} \, r_h^{D-2}}{4} \left(1 + \frac{2(D-2) }{(D-4)}\frac{\tilde \alpha}{r_h^2}\right)\,.
}
Here again, the sum of the Euclidean actions may be reduced to the sum of the entropies, which may be directly evaluated using the Girard-Newton formulas. In $D=6$, this was computed in \cite{Xu:2013zpa}. We have calculated the corrections in higher dimensions, which allows us to 
conjecture a closed form for even $D > 4$,
\eq{
\sum_h I_h =\frac{(-1)^{D/2}{\mathcal A}_{D-2}}{2} \sum_{n=0}^{\lfloor D/4 \rfloor} c_n \tilde\alpha^n L^{D-2-2n} 
\label{sumGB}
\,,}
where
\EQ{
&c_n = \frac{(-1)^n}{(2n)!!} \frac{\prod_{i=1}^{2n}(D+2-2i)}{\prod_{i=1}^{n} (D-2-2i)}
\,.
}
Note that we have not used the Smarr formula because the chemical potential conjugate to $\tilde \alpha$ contains a term proportional to the temperature \cite{Cai:2013qga}, which cancels with $\beta_h$ and the residue argument cannot be applied.
It would be an interesting problem to prove \eqref{sumGB}.
The series may be resummed using {\it Mathematica} as
\eq{
\sum_h I_h =\frac{(-1)^{D/2}{\mathcal A}_{D-2}}{2}\, L^{D-2} \,_2F_1\left(\frac{1}{2}-\frac{D}{4},-\frac{D}{4};2-\frac{D}{2};\frac{4 \alpha }{L^2}\right) \,.
}

In flat spacetime, the sum of the inverse temperatures vanishes whereas the sum of the Euclidean actions is corrected as
\EQ{
\sum_h I_h = \frac{(-1)^{D/2}{\mathcal A}_{D-2}}{2}  \frac{D}{D-4} \tilde \alpha^{\frac{D-2}{2}} \,.
}
It may be obtained from the relation of power sums 
$\sum_h r_h^{k} = -\tilde \alpha \sum_h r_h^{k-2}$ that follows from applying the Girard-Newton formulas to \eqref{DeltaGB}

\section{Discussion}
In this work, we have rigorously derived a universal result that applies to a large class of black-hole spacetimes. Individually on each ``unphysical'' horizon, the thermodynamic quantities such as temperature and entropy may be negative or even complex and can only be interpreted formally at present. The point is is that it can be physically relevant when all the horizons are taken in account.

As we have pointed out in the introduction, in the scalar wave equation, which is an important probe of a black hole, all the inner roots, real or complex, has effect on the equation even if we only examine the outer region. This and our results suggest that in order to understand the properties of black holes completely, it is necessary to complexify the metrics and take into account of all the special points in the complex space.

For asymptotically (A)dS spacetimes, the dependence of the sum of the Euclidean actions only on $\Lambda$ is suggestive of the property of the global AdS vacuum. Indeed, consider a pure (A)dS spacetime, with the metric
\eq{
ds_D^2 = -\left(\frac{r^2}{L^2} +k\right) dt^2 + \left(\frac{r^2}{L^2} + k\right)^{-1} dr^2 + r^2\, d\Omega_{D-2,k}^2 \,,
}
where $d\Omega_{D-2,k}$ is the volume element on a space with constant curvature and $k=1,0,-1$ corresponds to a sphere, Ricci flat, and hyperbolic topology, respectively. 
$L^2$ is positive for AdS and negative for dS.
The generalized horizons are at $r_\pm =\pm \sqrt{-k L^2}$, with 
\eq{
\beta_\pm = \pm \frac{\sqrt{-kL^2}}{2\pi L^2}
\,.}
Since the vacua have zero mass, we conclude that
\EQ{
\sum_{h} I_h &= -\sum_h S_h\cr
 &= -\frac{{\mathcal A}_{D-2,k}}{4} \left[\left(\sqrt{-k L^2}\right)^{D-2} +\left(-\sqrt{-k L^2}\right)^{D-2}\right].
}
Here $\mathcal{A}_{D,k}$ is the volume of the constant-curvature space. For black holes, the topologies change, but the sum of the Euclidean actions over all the horizons remains the same. This indicates that the outer region of a black hole is insufficient to address the spacetime properties and a new symmetry may emerge to extend the metric into the complex regions \cite{Lu:2024cvx}.

\acknowledgments
This work is partly supported by the National Natural Science Foundation of China (NSFC) Grants No.~11935009 and No.~12375052.

\bibliography{bib}
\end{document}